An analysis of states in the phase space: the anharmonic oscillator

Sebastiano Tosto

sebastiano.tosto@enea.it
stosto@inwind.it

ENEA Casaccia, via Anguillarese 301,
00123 Roma, Italy

ABSTRACT

The paper introduces a simple quantum model to calculate in a general way allowed frequencies and energy levels of the anharmonic oscillator. The theoretical basis of the approach has been introduced in two early papers aimed to infer the properties of quantum systems exploiting the uncertainty principle only. For clarity the anharmonic oscillator is described having in mind the lattice oscillations of atoms/ions, yet quantum formalism of the model and approach have general character and can be extended to any oscillating system. The results show that the harmonic energy levels split into a complex system of energy levels dependent upon the number of anharmonic terms that characterize the oscillator.

1. Introduction

The anharmonic phenomena, well known in physics [1], cover a wide range of properties of practical and theoretical interest; e.g. in acoustics they account for large variations of sound velocity in solids [2], in optics for non-linear interaction of powerful light with lattice vibrations [3]. Moreover are known physical effects that lead to a behavior impossible in harmonic oscillators, like the "foldover effect" [4] and "superharmonic resonance" [5]; both are due to the dependency of the eigenfrequency of nonlinear oscillators on the amplitude and to the non-harmoniticity of the oscillations. In solid state physics, non-linear effects occur when atoms consisting of a positively charged nucleus surrounded by a cloud of electrons are subjected to an electric field; the displacement of nucleus and electrons causes an electric dipole moment, whose interaction with the applied field is linear for small field intensities only [6]. The present paper aims to propose a quantum mechanical approach to tackle the problem of non-harmonic oscillations in a general way, i.e. regardless of the particular issue of specific interest, and in line with the concepts introduced in two papers [7,8] concerning simple quantum systems, many-electron atoms/ions and diatomic molecules. The basic idea of these papers starts from a critical review of positions and momenta of interacting particles in a quantum system, where the dynamical variables are perturbed in a complex way by mutual interactions and change within appropriate ranges of values in agreement with boundary conditions like the minimum total energy. Consider for instance the hydrogenlike atoms. It is reasonable to regard radial momentum $p_\rho$ and distance $\rho$ between electron and nucleus as variables included within proper ranges of values; it is certainly possible to write $0 < \rho \leq \Delta\rho$ and $0 < p_\rho \leq \Delta p_\rho$ if $\Delta\rho$ and $\Delta p_\rho$ have arbitrary sizes, including even the chance of infinite sizes. The basic hypothesis of the quoted papers was that $\Delta\rho$ and $\Delta p_\rho$ have physical meaning of quantum uncertainty ranges, thus to be regarded according to the basic ideas of quantum statistics; hence

$$\Delta x \Delta p_x = n\hbar \qquad 1,1$$

with $n$ arbitrary integer. No hypothesis is necessary about $\Delta x$ and $\Delta p_x$, which are by definition arbitrary, unknown and unpredictable. Eq 1,1 was the unique assumption in [7,8] and does so also in the model proposed here. Despite the apparently agnostic character of eq 1,1, the results inferred in the quoted papers were in all cases completely analogous to that of the usual wave mechanics formalism; in particular it was found that the quantum numbers actually coincide with the numbers of allowed states in the phase space for the concerned systems. Eq 1,1 only is enough to give the classical Hamiltonian, $H_{cl}$, the physical meaning of quantum Hamiltonian, $H_q$; it simply requires considering the ranges of dynamical variables rather than the dynamical variables themselves, which are therefore disregarded since the beginning. For instance, in a one-dimensional problem like that of a mass constrained to oscillate along a fixed direction, it means that hold the positions

$$H_{cl}(x, p_x) \Rightarrow H_q(\Delta x, \Delta p_x) \Rightarrow H_q(\Delta p_x, n) \qquad 1,2$$

The uncertainty is regarded in this way as fundamental principle of nature rather than as mere consequence of commutation rules of quantum operators. The case of the harmonic oscillator, already introduced in [7], has central importance here; its quantum formulation according to eq 1,1 and positions 1,2 is so short and simple that it is sketched in the next section 2 to make the present paper clearer and self-consistent. The next section aims also to show how the concepts so far introduced enable the quantum approach. For clarity the anharmonic oscillator is regarded in section 3 having in mind the lattice oscillations of atoms/ions, yet through a very general model. The discussion on the results of the model and the conclusion are reported in sections 4 and 5.

2. The harmonic oscillator.

With the positions 1,2, the classical energy equation $E = p^2/2m + k_{har}(x-x_o)^2/2$ of the oscillating mass around the equilibrium position $x_o$ reads $\Delta E = \Delta p^2/2m + k_{har}\Delta x^2/2$, having omitted for simplicity the subscript $x$; owing to eq 1,1, $E = E(\Delta p, n)$ is now because of $n$ a random quantity within an energy range $\Delta E$ that corresponds to local uncertainty of dynamical variables within $\Delta x$ and $\Delta p$. Both these latter and $\Delta E$ are assumed positive by definition. Then, one finds

$$\Delta E = \frac{\Delta p^2}{2m} + \frac{m(n\hbar\omega_{har})^2/2}{\Delta p^2} \qquad \omega_{har}^2 = \frac{k_{har}}{m} \qquad 2,1$$

Eq 2,1 has a minimum as a function of $\Delta p$, i.e.

$$\Delta p_{min} = \sqrt{mn\hbar\omega_{har}} \qquad \Delta E_{min} = n\hbar\omega_{har} \qquad 2,2$$

being now $n$ the number of vibrational states. Although for $n=0$ there are no vibrational states, the necessity that $\Delta p \neq 0$ compels $\Delta E \neq 0$ and thus $\Delta E_0 = \Delta p_0^2/2m \neq 0$ with $\Delta p_0 = \Delta p_{min}(n=0)$. In this particular case, the problem reduces to that of a free particle in the box, i.e. $\Delta p_0$ is related to the zero point energy. This requires $\Delta p_0 = \Delta p_{min}(n=1)$, because the minimum quantum uncertainty of $\Delta p$ can be nothing else but that of $\Delta p_{min}$ for $n=1$. The numerical correspondence between non-vibrational momentum range, $\Delta p_0$, and first vibrational momentum range, $\Delta p_{min}(n=1)$, means that at the zero point energy state the mass $m$ is delocalized in a space range, $\Delta x_0 = \Delta x(n=0)$, equal to that, $\Delta x(n=1)$, pertinent to the lowest vibrational state. In other words, the oscillation amplitude at the ground energy level is the same as the delocalization range size of the particle with zero point energy only. Hence $\Delta p_0 = \sqrt{m\hbar\omega_{har}}$ defines $E_0 = \Delta p_0^2/2m = \hbar\omega/2$. The minimum of $\Delta E$ must be $\Delta E_{min} = E_{min} - \hbar\omega_{har}/2$; then, regarding $E_{min} = E_{har}$ as the harmonic energy level, the known result

$$E_{har} = n\hbar\omega_{har} + \hbar\omega_{har}/2 \qquad 2,3$$

is obtained considering uncertainty ranges of eq 1,1 only, without any further hypothesis. Note that $\Delta p^2/2m = \omega_{har}^2 mn^2\hbar^2/2\Delta p^2 = n\hbar\omega_{har}/2$ with $\Delta p = \Delta p_{har}$, in agreement with the virial theorem; $E_{min}$ is given by the sum of kinetic and potential terms, whereas the zero point term has kinetic character only. Also note in this respect that $\Delta p_{min}$ and $\Delta p_0$ are merely particular range sizes, among all the ones allowed in principle, fulfilling the condition of minimum $E_{min}$ and $E_0$. These results do not contradict the complete arbitrariness of $\Delta p$ and $\Delta x$, since in principle there is no compelling reason to regard the particular ranges of eqs 2,2 in a different way with respect to all the other ones allowed by eq 1,1; rather the results merely show the preferential propensity of nature for the states of minimum energy. In effect it is not surprising that the energy calculated with extremal values of dynamical variables in the ranges of eq 2,1 does not coincide, in general, with the most probable energy. In conclusion, this example highlights that the physical properties of a quantum system can be inferred without solving any wave equation simply replacing the local dynamical variables with the respective quantum uncertainty ranges: the key problem becomes then that of counting correctly case by case the appropriate number of allowed states, as shown in [7,8] for more complex quantum systems. It appears that, once accepting the eq 1,1 and the positions 1,2, have actual physical meaning the uncertainty ranges rather than the dynamical variables themselves; these latter are considered here random, unknown and unpredictable within the respective ranges and thus are disregarded since the beginning when formulating the physical problem. Just this is the essence of eq 2,1. Since the present approach gives sensible results for harmonic oscillations, there is no reason to exclude that the same holds for anharmonic oscillations as well. The next paragraph aims to generalize the kind of approach just introduced to the case of anharmonic oscillations.

3. The anharmonic oscillator.
The classical Hamiltonian reads now

$$E = p^2/2m + \sum_{i=2}^{N} a'_i(n\hbar)^{-i}(x-x_o)^i \qquad 3,1$$

being $N$ the arbitrary number of terms of the series including quadratic and anharmonic terms and $a'_i$ proper coefficients. The values of these coefficients, assumed known, distinctively characterize the specific kind of oscillating system. The signs of $a'_3$ and $a'_4$ are taken here negative [9]; the former expresses the asymmetry of the mutual repulsion between atoms or ions, e.g. in a metallic lattice, the latter describes the softening of the vibration at large amplitudes. The higher order terms allow to describe these effects in a more general way, so their sign and values must agree with the idea that the global consequence of anharmonicity is to lower the potential energy of oscillation; indeed the potential energy reads $a'_2(x-x_o)^2 f(x)$, i.e. it consists of a quadratic term with $x$-dependent correction factor $f(x) = 1 + \sum_{i=3}^{N}(a'_i/a'_2)(x-x_o)^{i-2} < 1$. By analogy with the harmonic case, the coefficient of the quadratic term, anyway related to the force constant $k_{an}$, is reasonably expected to have still the form $m(n\hbar\omega_{an})^2/2$ with oscillation frequency defined now by $\omega_{an}^2 = k_{an}/m$. Moreover the dependence of this term on $\omega_{an}$ suggests that in general $a'_i = a'_i(\omega_{an})$ are to be expected as well. The following discussion aims to guess this dependence and the relationship between $\omega_{an}$ and $\omega_{har}$ through the same approach shown previously; so, as done in section 2, we aim to calculate $\Delta E_{\min}$ and infer next the anharmonic vibrational levels $E_{an}$ and zero point energy $E_0$, being clearly $\Delta E_{\min} = E_{\min} - E_0$ and $E_{an} = E_{\min}$. According to the position 1,2 and eq 1,1, the quantum energy equation corresponding to eq 3,1 reads

$$\Delta E = \frac{\Delta p^2}{2m} + \sum_{i=2}^{N} \frac{a'_i}{\Delta p^i} \qquad 3,2$$

This equation, minimized with respect to the range $\Delta p$, yields

$$\Delta p_{\min} = m \sum_{i=2}^{N} i a'_i \Delta p_{\min}^{-(i+1)} \qquad \Delta E_{\min} = \Delta E(\Delta p_{\min}) \qquad \Delta p_{\min} = \Delta p_{\min}(\omega_{an}) \qquad 3,3$$

For assigned coefficients $a'_i$, the first equation admits in general $N+2$ solutions $\Delta p_{\min}$, some of which can be however imaginary. Being the momentum uncertainty range $\Delta p_{\min}$ positive by definition, let $I' \leq N+2$ be the number of positive real roots; so $I'$ possible values of $\Delta p_{\min}$ describe the allowed momentum ranges of the oscillating particle that fulfil the minimum condition. A further limitation to these values is that the series must converge. Disregard also the values of $\Delta p_{\min}$ that with the given $a'_i$ possibly do not fulfil the inequality $\left|(i+1)a'_{i+1}\Delta p_{\min}^{-(i+2)}\right| << \left|ia'_i\Delta p_{\min}^{-(i+1)}\right|$ inferred from eq 3,2, i.e.

$$\left|a'_{i+1}\right| << \left|a'_i\Delta p_{\min}\right| \qquad 3,4$$

Then $I \leq I'$ is the number of real roots of physical interest to be considered in the following. Trivial manipulations of eq 3,2 to eliminate $m$ with the help of eq 3,3 yield

$$\Delta E = \frac{1}{2}\left(\frac{\Delta p}{\Delta p_{\min}}\right)^2 \sum_{i=2}^{N}\frac{ia'_i}{\Delta p_{\min}^i} + \sum_{i=2}^{N}\frac{a'_i}{\Delta p^i} \qquad 3,5$$

To extract the allowed physical information from this equation one should minimize with respect to $\Delta p$ and then proceed as shown in the harmonic case. Actually this minimum condition has been already exploited to infer eq 3,3, which suggests that eq 3,5 should not need being minimized once more. To understand this point replace $\Delta p$ with $\Delta p_{\min}$ in eq 3,5 and consider first the resulting equation $\Delta E(\Delta p_{\min}) = \sum_{i=2}^{N}(1+i/2)a'_i\Delta p_{\min}^{-1}$ in the harmonic case; then $N = 2$, i.e. $a_{i>2} = 0$, yields $3a'_2\Delta p_{\min}^{-2}/2$. By comparison with eq 2,1 this result takes a more familiar form replacing $a'_2$ with $a_2\Delta p_{\min}^4/m$, where $a_2$ is a dimensionless proportionality coefficient linking $a'_2$ and $\Delta p_{\min}$; in this way one obtains $\Delta E(\Delta p_{\min}) = 3a_2\Delta p_{\min}^2/2m$, which has the same form of eqs 2,2 a proportionality factor apart. As expected, an immediate connection with the harmonic case is possible uniquely on the basis of the condition 3,3 without introducing explicitly neither $\omega_{har}$ nor the equations of $\Delta p_{har}$ and $\Delta E_{har}$. Express thus in general the coefficients $a'_i$ as a function of $\Delta p_{\min}$ as follows

$$a'_i = \frac{\Delta p_{\min}^{i+2}}{m}a_i \qquad \sum_{i=2}^{N} ia_i = 1 \qquad 1 \leq j \leq I \qquad 3,6$$

where $a_i$ are new constants that fulfil the boundary condition expressed by the second equation, straightforward consequence of eq 3,3. Note that $a'_i$ are uniquely defined for the specific oscillating system, whereas the appropriate notation of the various $a_i$ should be $a_i^{(j)}$ to emphasize that a set of these coefficients is defined by each solution $\Delta p_{\min}^{(j)}$ of physical interest calculated through eq 3,3. This would also compel indicating in eq 3,5 $\Delta E^{(j)}$ and then $\Delta E_{\min}^{(j)}$. To simplify the notations the superscript $(j)$ will be omitted, stressing however once for all that if $N > 2$ then eq 3,5 actually represents anyone among $I$ admissible equations. Replacing $a'_i$ into the energy equation 3,5, one finds $\Delta E = \left((\Delta p/\Delta p_m)^2/2 + \sum_{i=2}^{N} a_i(\Delta p_m/\Delta p)^i\right)\Delta p_{\min}^2/m$; this suggests putting

$$q\frac{\Delta E}{\Delta E_{\min}} = \frac{1}{2}\left(\frac{\Delta p}{\Delta p_{\min}}\right)^2 + \sum_{i=2}^{N} a_i\left(\frac{\Delta p_{\min}}{\Delta p}\right)^i \qquad \Delta E_{\min} = q\frac{\Delta p_{\min}^2}{m} \qquad a_2 = \frac{1}{2}\left(1 - \sum_{i=3}^{N} ia_i\right) \qquad 3,7$$

The proportionality factor $q$ aims to fulfil the reasonable condition $\Delta E = \Delta E_{\min}$ for $\Delta p = \Delta p_{\min}$ and express in a general way the expected link between $\Delta E_{\min}$ and $\Delta p_{\min}^2/m$. Trivial calculations yield

$$q = 1 + \sum_{i=3}^{N}(1 - i/2)a_i$$

Of course $q$ must be intended here as $q^{(j)}$ likewise as $a_i^{(j)}$. Whatever $a_i$ might be, eq 3,7 does not need being minimized; it simply expresses as a function of $\Delta p / \Delta p_{min}$ the energy deviation from the harmonic condition for assigned values of the coefficients $a'_{i \geq 2} \neq 0$. Eq 3,7 and $a_2$ are uniquely defined in the particular case $a_{i>2} = 0$ only, which corresponds to $q = 1$ as well. Moreover the form of the second equation, analogous to that of eqs 2,2, suggests that $\Delta p_{min}$ and $\Delta E_{min}$ must be also equal or proportional to the respective harmonic quantities $\Delta p_{har}$ and $\Delta E_{har}$. So putting in general $\Delta E_{min} = w^2 \Delta E_{har}$ and $\Delta p_{min} = w \Delta p_{har}$, with $w$ proportionality factor, one finds

$$\frac{q}{w^2} \frac{\Delta E}{\Delta E_{har}} = \frac{1}{2w^2} \left( \frac{\Delta p}{\Delta p_{har}} \right)^2 + \sum_{i=2}^{N} a_i w^i \left( \frac{\Delta p_{har}}{\Delta p} \right)^i \qquad \omega_{an} = w^2 \omega_{har} \qquad 3,8$$

Likewise $q$, also $w$ must be intended in general as $w^{(j)}$. So eqs 3,8 define $I$ anharmonic frequencies $\omega_{an}^{(j)} \neq \omega_{har}$, here designated shortly $\omega_{an}$, corresponding to the unique harmonic frequency $\omega_{har}$; i.e. the various $\Delta E_{min}$ describe the splitting of each $n$-th vibrational energy level $n\hbar\omega_{har}$. The anharmonic potential of eq 3,8 is expected to depend upon $\omega_{an}$ through the dimensionless coefficients $a_i$, by analogy with the dependence of the harmonic term upon $\omega_{har}^2$. Thus, to complete the task of the present section it is necessary: (i) to define the factor $w$ of eq 3,8; (ii) to highlight the analytical form of the functions $a_i = a_i(\omega_{an})$; (iii) to express the potential energy of equation 3,8 as a function of $\omega_{an}$ through these coefficients. Rewrite to this purpose the coefficients of eq 3,2 as shown in following series

$$q \Delta E = \frac{1}{2} \frac{\Delta p^2}{m} + \sum_{i=2}^{N} a''_i \frac{m^{i/2}(n\hbar\omega_{an})^{i/2+1}}{\Delta p^i} \qquad 3,9$$

where the powers of $n\hbar\omega_{an}$ and $m$ have been determined by dimensional consistency of the various terms with both $\Delta E$ and $\Delta p^i$. Minimizing with respect to $\Delta p$ and equating to zero, one finds

$$R_E = \frac{1}{2} R_p^2 + \sum_{i=2}^{N} a''_i R_p^{-i} \qquad R_E = q \frac{\Delta E}{\Delta E_{har}} \qquad R_p = \frac{\Delta p}{\Delta p_{har}} \qquad a''_i = a_i w^{i+2} \qquad 3,10$$

With the coefficients $a''_i$ and $a_i$ linked by the last position, eq 3,10 is identical to eq 3,8; this consistency supports therefore the positions of both eqs 3,6 and 3,9. To specify $w$ put first $N = 2$ in eq 3,8; minimizing $R_p^2 / 2w^2 + a_2 w^2 / R_p^2$ with respect to $R_p$ yields $R_p^4 = 2a_2 w^4$. Since the minimum of $R_p$ can be nothing else but 1 by definition, $w = (2a_2)^{-1/4}$ yields $w = 1$, whereas in this particular case $a_2 = 1/2$. Also eq 3,8 is thus uniquely defined for $a_{i>2} = 0$ only. Note that the coefficient of the quadratic term of eq 3,9 reads $a''_2 m(n\hbar\omega_{an})^2$; if the result $w = (2a_2)^{-1/4}$ previously obtained for $N = 2$ still holds for any $N$ with $a_2$ given now by the last eq 3,7, then $a''_2 = a_2 w^4$ yields $a''_2 = 1/2$ and thus the expected form $m(n\hbar\omega_{an})^2 / 2$ formerly quoted whatever $a_{i>2}$ might be. This consideration encourages one to conclude with the help of eq 3,7

$$w^2 = (2a_2)^{-1/2} = \left(1 - \sum_{i=3}^{N} i a_i\right)^{-1/2} \qquad a_i'' = a_i \left(1 - \sum_{i=3}^{N} i a_i\right)^{-i/4 - 1/2}$$

Replacing $a_i''$ in eq 3,9 one finds

$$\Delta E = \frac{1}{2q} \frac{\Delta p^2}{m} + \sum_{i=2}^{N} q^{-1} a_i \left(1 - \sum_{i=3}^{N} i a_i\right)^{-\frac{3}{4}(i+2)} \frac{m^{i/2} (n\hbar \omega_{har})^{i/2+1}}{\Delta p^i} \qquad 3,11$$

This is the generalization of eq 2,1 when $a'_{i>2} \neq 0$; the positions so far introduced link eq 3,2 with the harmonic case. Moreover eq 3,8 yields

$$\omega_{an} = \left(1 - \sum_{i=3}^{N} i a_i\right)^{-1/2} \omega_{har} \qquad 3,12$$

With the given choice of $w^2$, therefore, $a_{i \geq 3} = 0$ yield not only $\omega_{an} = \omega_{har}$ but also $\Delta p_{min} = \Delta p_{har}$ and $\Delta E_{min} = \Delta E_{har}$. Hence

$$\Delta E_{min} = n\hbar \omega_{an} = \left(1 - \sum_{i=3}^{N} i a_i\right)^{-1/2} n\hbar \omega_{har} \qquad \Delta p_{min} = \sqrt{mn\hbar \omega_{an}} = \left(1 - \sum_{i=3}^{N} i a_i\right)^{-1/4} \sqrt{mn\hbar \omega_{har}} \qquad 3,13$$

As concerns the zero point energy $E_0$ hold the considerations of the previous section, i.e. $\Delta E_{min} = E_{min} - E_0$; moreover also now for $n=0$ the minimum of eq 3,11 reduces to $\Delta p_0^2 / 2qm$, with $\Delta p_0^2 = \Delta p_{min}^2(n=0)$. As explained before, even in lack of vibrational states $\Delta p_{min} \neq 0$ compels putting $\Delta p_0 = \Delta p_{min}^{(0)}(n=1)$ by virtue of eq 3,13 so that $E_0 = \left(1 - \sum_{i=3}^{N} i a_i^{(0)}\right)^{-1/2} \hbar \omega_{har} / 2q$; since in general are allowed several values of $\Delta p_{min}$, the notation emphasizes that one must consider here the set of values of $a_i^{(j)}$ corresponding to the smallest among the various $\Delta p_{min}^{(j)}$. In conclusion, since the anharmonic energy and momentum must correspond to the respective $\Delta E_{min}$ and $\Delta p_{min}$, it is possible to summarize the previous results, with full notation for clarity, as follows

$$E_{an}^{(j)} = \left(1 - \sum_{i=3}^{N} i a_i^{(j)}\right)^{-1/2} n^{(j)} \hbar \omega_{har} + \frac{1}{2q}\left(1 - \sum_{i=3}^{N} i a_i^{(0)}\right)^{-1/2} \hbar \omega_{har} \qquad \Delta p_{an}^{(j)} = \left(1 - \sum_{i=3}^{N} i a_i^{(j)}\right)^{-1/4} \sqrt{mn^{(j)} \hbar \omega_{har}}$$

$$\omega_{an}^{(j)} = \left(1 - \sum_{i=3}^{N} i a_i^{(j)}\right)^{-1/2} \omega_{har} \qquad a_i^{(j)} = \frac{m a_i'}{\left(\Delta p_{min}^{(j)}\right)^{i+2}} \qquad 1 \leq j \leq I \qquad 3,14$$

4. Discussion.

The strategy of the papers [7,8] to exploit via eq 1,1 the classical Hamiltonians of the system of interest was outlined in section 2 and then extended in section 3 to the anharmonic case. The first task of the discussion aims to clarify the classical and quantum ways to regard the harmonic and anharmonic oscillation. The classical potential energy of eq 3,1, $U_{cl} = U_{cl}(x - x_o)$, concerns a withholding force progressively increasing as a function of $x - x_o$ while the oscillation turns gradually from harmonic into anharmonic behaviour. Moreover if momentum and position of $m$ are both exactly known, $U_{cl}$ can be defined with arbitrary accuracy simply increasing the number of terms of the series. This description is clearly inadequate for the potential energy, $U_q = U_q(\Delta p, n)$, of the quantum eq 3,2; in principle the exact elongation of $m$ with respect to the rest position and the corresponding momentum are not jointly specifiable, i.e. the limit $\Delta x \to 0$ could not be described by finite values of $\Delta p_{min}$. Indeed $\Delta p_{min} \to \infty$ compels $\Delta p \to \infty$ that yields $\Delta E = \Delta p^2 / 2m$ regardless of $a'_i$; this limit corresponds to the classical case of a free particle in a one-dimensional box, of no interest here, rather than to the harmonic limit expected for $\Delta x \to 0$. Eventually the quantum uncertainty compels regarding in a different way also the number of terms of $U_{cl}$ and of $U_q$: in the former case $N$ is in principle arbitrary, being significant its ability to provide a description as detailed as possible of the local state of motion of $m$, in the latter case does not, being instead significant its ability to introduce the allowed physical information into the system. If for instance the model aims to describe softening and asymmetry effects only, then are justified terms like $\Delta x^i$ with powers and signs pertinent to these effects only [9]. Solving eq 3,1 requires exploiting the functional relationship $U_{cl}$ upon $\Delta x$ through numerical methods, solving eq 3,2 requires instead a different reasoning because the anharmonic effects inherent the various $\Delta x^i$ are related to the respective $\Delta p^{-i}$ through eq 1,1 only: the previous results show that a general physical principle, the minimum energy, is enough to this purpose. According to the classical eq 3,1 the harmonicity requires $a'_{i \geq 3} \Delta x^i << a'_2 \Delta x^2$ in agreement with the convergence condition 3,4; the quantum eq 3,2 requires $a'_{i \geq 3} \Delta p^{-i} << a'_2 \Delta p^{-2}$, which is still a statement of "small" oscillation amplitudes since $a'_i \Delta p^{-i} \propto a'_i \Delta x^i$. Both definitions are thus equivalent, yet the latter is more interesting because it involves eq 1,1 and allows further considerations on the classical and quantum concepts of harmonicity. Eq 3,4 and the first eq 3,3 yield

$$a'_{i \geq 3} \Delta p^{-i} << a'_2 \Delta p^{-2} \quad \Rightarrow \quad a_i \left( \frac{\Delta p_{min}}{\Delta p} \right)^i << a_2 \left( \frac{\Delta p_{min}}{\Delta p} \right)^2 \qquad i \geq 3$$

Noting that $\Delta p$ is arbitrary by assumption and that $\Delta p_{min} \leq \Delta p$ by definition, it turns out that the second inequality can be merely fulfilled by $\Delta p / \Delta p_{min} >> 1$ regardless of the values of the ratios $a_2 / a_i$ and $a'_2 / a'_i$. Since in principle $a'_i$ only are required to fulfil the convergence condition 3,4 whereas $a_i$ do not, because their values are consequently defined in the successive eq 3,6 only, the conclusion is that small oscillation amplitudes do not require necessarily vanishing $a_{i>2}$. According to eq 3,12, however, just these latter define $w$ that in turn control $\omega_{an}$ and thus the splitting of energy levels. The fact that in general $w \equiv w^{(j)} \neq 1$ even for small oscillations supports the idea that the quantum harmonicity is a particular case, but not a limit case, of the quantum anharmonicity; in other words, an oscillating quantum system does not change gradually from harmonic to anharmonic behaviour. This conclusion is confirmed also considering the dependence of the constants $w$ on $a_i$. In eq 3,6 large values of $\Delta p_{min}$ entail small $a_i$ and thus $w$ such that the corresponding allowed frequencies $\omega_{an}$ are expected to have values similar to $\omega_{har}$; the contrary holds for small values of $\Delta p_{min}$, to which correspond larger values of $w$ and therefore larger gaps $\omega_{an} - \omega_{har}$. Hence, when considering the totality of allowed frequencies consistent with the different sizes of all ranges $\Delta p_{\min}$, even small values of $a'_i$ classically compatible with the harmonic condition entail anyway relevant splitting and gap of energy levels with respect to $\omega_{har}$ typical of the anharmonicity; otherwise stated, the quantum harmonicity requires $a'_{i \geq 3} = 0$ exactly. The harmonic ground level is a reference energy rather than an attainable limit energy because fails the classical expectation of anharmonic frequencies progressively deviating from $\omega_{har}$ along with $a'_i$; the last eq 3,7 shows indeed that even the first quadratic coefficient $a_2$ of potential energy differs from the corresponding harmonic coefficient unless $a_{i \geq 3} = 0$. It is also significant the fact that the unique $\omega_{har}$, classically defined in eq 2,1 through the force constant $k_{har}$ of Hooke law only, never corresponds to a unique $\omega_{an}$ whatever $a'_{i \geq 3} \neq 0$ might be; this latter, although formally introduced in the early eq 3,3 as $\omega_{an}^2 = k_{an}/m$, has quantum character after being subsequently redefined by eq 3,8 through the multiplicity of values of $w$. It is however worth noting in this respect a further chance to define the oscillation frequency in a mere quantum way through an uncertainty equation having a form seemingly different but conceptually equivalent to eq 1,1. Introduce the time range $\Delta t$ necessary to displace $m$ by $\Delta x$ with finite average velocity $\mathbf{v}$; defining then $\Delta t = \Delta x / \mathrm{v}_x$ and $\Delta E = \Delta p \mathrm{v}_x$, eq 1,1 takes a form that introduces new dynamical variables $t$ and $E$ having random, unpredictable and unknown values within the respective uncertainty ranges defined by the same $n\hbar$. Of course $\Delta t$ and $\Delta E$ are completely arbitrary, as they must be, likewise $\Delta x$ and $\mathbf{v}$. Thus, with the constrain of equal $n$, eq 1,1 reads also

$$\Delta E \Delta t = n\hbar \qquad \Delta t = t - t_o \qquad 5,1$$

Eq 5,1 is not a trivial copy of eq 1,1: it introduces new information through $\mathbf{v}$ and shows that during successive time steps $\Delta t$ the energy ranges $\Delta E$ change randomly and unpredictably depending on $n$. Of course the eq 1,1 could have been inferred itself in the same way from eq 5,1, i.e. regarding this latter as the fundamental statement. Relating eqs 1,1 and 5,1 via the same arbitrary integer $n$, whatever it might be, means describing the oscillation of $m$ through energy and time uncertainty ranges. To show the consequences this assertion, consider that $1/\Delta t$ has in general physical dimensions of frequency; then eq 5,1 can be rewritten as $\Delta E_n = n\hbar \omega^\S$, being $\omega^\S$ a function somehow related to any frequency $\omega$. If in particular $\omega^\S$ is specified to be just the previous frequency $\omega_{har}$, whatever the value of this latter might be, eq 5,1 reads

$$\Delta E_n = n\hbar \omega_{har} \qquad 5,2$$

The notation emphasizes that the particular case $\omega^\S \equiv \omega_{har}$ enables a direct conceptual link with eq 2,3, i.e. it concerns the harmonicity; having found that $n$ is according to eq 1,1 the number of vibrational states of the oscillator and $n\hbar\omega_{har}$ their energy levels, then without need of minimizing anything one infers that $\Delta E_n$ is again the energy gap between the $n$-th excited state of the harmonic oscillator and its ground state of zero point energy; the condition of minimum energy and $\Delta p_{\min}$ are now replaced by the specific meaning of $\Delta t$. This conclusion shows that a particular property of the oscillating system is correlated to a particular property of the uncertainty ranges, thus confirming the actual physical meaning of these latter. So $E_n$ falling within $\Delta E_n$ are still now random, unpredictable and unknown because of $n$. While $\omega_{har}$ was formerly defined by the formal position 2,1, now eq 5,1 reveals its actual quantum meaning due to its direct link with the time uncertainty $\Delta t$. This last result is significant for the present discussion: it justifies the different outcomes of the quantum approach with respect to the classical expectation in terms of uncertainty about the dynamical variables of $m$ only; thus, as shown in [7,8], this result disregards any phenomenological/classical hint to describe the system. In other words, instead of thinking to a withholding spring bound to a mass moving back and forth, the oscillation can be imagined in a more abstract way. It is enough to introduce an arbitrary energy range $\Delta E_n$ to which corresponds a respective quantum frequency $1/\Delta t_n$; then the form of eq 1,1 is suitable to introduce an appropriate potential energy with elongation extent described by a unique quadratic term or by a series of terms, whose coefficients are respectively expressed as a function of $\omega_{har}$ or $\omega_{an}$ like in eqs 2,1 or 3,9. The worth of this conclusion is due to the generality of the resulting concept of oscillation, which skips any information on actual kind of motion of $m$, particular property of the oscillating mass, specific nature of the withholding force and hypothesis on the allowed range of frequencies. The previous results highlight the link of the allowed frequencies to the terms of $U_q$, see in particular the remarks about eqs 3,5 and 3,12. A consequence of this point of view is that replacing $U_{cl}$ with $U_q$ compels the existence of several momentum uncertainty ranges $\Delta p_{min}$ and thus of as many $\omega_{an}$ even when one would expect a mere perturbation of the unique $\omega_{har}$: the physical information provided by the quadratic term only is uniquely defined, instead the various values of $\Delta p_{min}$ and $\omega_{an}$ for $N > 2$ in eq 3,2 reveal according to the last eq 3,7 multiple anharmonic effects that influence also the quadratic term. The quantum uncertainty is therefore crucial in describing the oscillation. For instance let us show that, at least for certain frequencies, the anharmonic oscillator appears to be a system intrinsically unstable. Let $i$ be the index of any high order term of the series such that $a'_i/\Delta p^i << a'_2/\Delta p^2$ is true by definition because of the convergence condition; so $a'_i/\Delta p^i$ represents a small contribution to the total energy of oscillation. Let $\delta a'_i/\Delta p^i$ be its value altered by the change of the coefficient $a_i$ because of an external perturbation acting on the oscillator; if for instance an impurity diffuses through the lattice in proximity of the given oscillating atom/ion, the stress field around this impurity or its possible charge field reasonably modify the local repulsion between atoms/ions or the softening effects at large oscillation amplitudes, as a consequence of which the anharmonic coefficients $a'_3$ and/or $a'_4$ are expected to change. Let us exemplify any perturbation like this through a suitable change of some $a'_i$ of the $i$-th energy terms in eq 3,2; here however we consider for simplicity one term only to describe the local effect. The proof that some $\Delta p_{min}$ and resulting $\Delta E_{min}$ are strongly affected even by a very small change of any $a'_{i>2}$ is easy in the particular case where the series describing the potential energy converges very quickly. Differentiating eq 3,6 one finds

$$\delta a'_i = a'_i \left( (i+2) \frac{\delta \Delta p_{min}}{\Delta p_{min}} + \frac{\delta a_i}{a_i} \right)$$

Fix the value of $\delta a'_i$; if the local perturbation of the lattice affects $a'_i$ in such a way that $\delta a'_i \gg a'_i$, i.e. it alters significantly $a_i$, then the quantity in parenthesis is very large. If this happens while holds for $\delta a'_i$ also the condition $\delta a'_i / \Delta p^i \ll a'_2 / \Delta p^2$, still possible because no hypothesis has been made on the strength of the perturbation, then considering that the quadratic term provides the most essential contribution to the total potential energy the result is: even a small perturbation $\delta a'_i / \Delta p^i$ of the whole oscillation energy is able to change significantly both $\Delta p_{min}$ and $a_i$ that define $\omega_{an}$, see eqs 3,12 to 3,14. The altered size of the range $\Delta p_{min}$, actually verified by preliminary numerical simulations carried out with arbitrary coefficients $a'_i$ matching the aforesaid assumptions, means in particular that the whole energy of the system admits not only the change of $\omega_{an}$ allowed to the oscillator but also a larger range of corresponding momenta $p_{min}$ allowed to $m$; this does not exclude even the chance of chaotic motion related to a random sequence of values $\omega_{an}$ during a weak perturbation transient of the diffusing impurity. The reason of such instability rests once again on the different way of regarding the oscillation amplitudes in classical and quantum physics. The former admits the limit $\Delta x \to 0$ regardless of $\Delta p$, the latter does not; so the quantum oscillation range of physical interest cannot be arbitrarily small or change arbitrarily without violating the crucial condition of minimum energy. Indeed the oscillation range sizes corresponding to the vibrational levels are quantized themselves

$$\Delta x_{min} = \sqrt{\frac{n\hbar}{\omega_{an} m}} \qquad \Delta x_0 = \sqrt{\frac{\hbar}{\omega_{an}^{(0)} m}}$$

At this point it is worth remembering what has been previously emphasized, i.e. that the sizes of the ranges $\Delta x$ and $\Delta p$ are unspecified and indefinable; $\Delta x_{min}$ and $\Delta p_{min}$ are merely particular values showing the propensity of nature to fulfil the condition of minimum energy, however without contradicting the assumption that the uncertainty ranges are in principle completely arbitrary. So oscillation ranges that do not fulfil the former condition are certainly possible but unstable because of mere quantum reasons, i.e. they do not correspond to momentum range sizes that minimize the oscillation energy levels. This conclusion is important because its validity follows uniquely from the assumption of convergence of the potential series only, i.e. it concerns a realistic condition effectively possible for the oscillator rather than an unusual and improbable limit case. Also, this result holds whatever the origin of the anharmonicity might be and confirms the physical diversity of harmonic and anharmonic quantum systems. Note however that the former is actually an ideal abstraction only; what can be expected in practice is a strong or weak anharmonicity, unless some specific physical reason requires just a potential energy with quadratic term only. So the results of the present approach should be regarded as the realistic behaviour of any oscillating system, rather than a sophisticated improvement of the naïve harmonic behaviour; now this latter appears thus in general reductive and incomplete, rather than merely approximate. Yet eq 3,14 shows that the zero point energy is formally analogous in both cases, a numerical difference apart: the only difference between the harmonic and anharmonic cases is that instead of considering the unique $\hbar\omega_{har}/2$ one must select the smallest $\omega_{an}^{(j)}$ to calculate $\hbar\omega_{an}^{(0)}/2$. Note eventually that easy considerations allow to generalize the concept of perturbed oscillator in the conceptual frame of the present model. So far the present approach aimed to introduce the terms $a_3$ and $a_4$ to account for the anarmonicity, so that eqs 3,2 to 3,14 tacitly assume an isolated oscillating system. Simple considerations however allow to further generalize the physical meaning of eq 3,2 taking advantage of the fact that the present model works with a number of high order terms in principle arbitrary. In particular coefficients and number of terms could be exploited to describe even an oscillating system perturbed by an external force, for instance due to the interaction with other oscillators; indeed this force can be certainly described as a series development having the form $\sum a_i'' \Delta x^i$ if it is related, in the most general case non-linearly, to the displacement extent of the oscillating mass. So, whatever the nature of the perturbation might be, this means that the potential energy of the system changes by an additional amount $-\sum a_i''(1+i)^{-1} \Delta x^{i+1}$ to be summed up with the corresponding terms of eq 3,1. In any case, however, adding an arbitrary number of such energy terms to those intrinsically characterizing the oscillator does not change in principle the approach so far exposed, except of course the numerical value of the various $a_i$ of eq 3,8, which are now replaced by the sum $a_i'' + a_i'$ for each $i$-th power of oscillation elongation. So nothing hinders to regard the energy range $\Delta E_{an}$ of this equation as $\Delta E_{an+pert}$ still normalized to that of an isolated harmonic oscillator; it is enough that the coefficients $a_i'$ up to the $N$-th order are still known, i.e. defined by the particular kind of oscillating system and external perturbation, yet without necessarily assuming any constrain on their signs, now determined by the sum of both effects. Even in the case where the force is described by terms like $\alpha'/\Delta x^i$ one would find an equation like 3,2 containing however terms like $a_k' \Delta p^k$ with $k > 2$. Also in this case, however, minimizing with respect to $\Delta p$ would yield an appropriate number of roots $\Delta p_{min}$ and thus conclusions in principle completely analogous to that previously carried out. In the present case holds therefore the following position

$$\omega_{an+pert} \gtreqless w^2 \omega_{har}$$

As expected, the previous scheme of vibrational levels is modified the external perturbation that affects $w$. This last result confirms the very general character of the way to describe any oscillating system simply with the help of the fundamental eq 1,1.

5. Conclusion.
The computational scheme introduced in the present paper is very simple: the most important achievements hitherto exposed do not require hard numerical calculations, but are consequence of general considerations on basic concepts of quantum mechanics. The general character of the approach, e.g. due to the arbitrary number $N$ of anharmonic terms, and the possibility of extension to the case of a perturbed oscillator, propose the model as a useful tool in a broad variety of physical problems.